\documentclass[11pt]{article}

\usepackage{amsmath}
\usepackage{amssymb}
\usepackage{braket}
\usepackage{graphicx}
\usepackage{authblk}
\usepackage[margin=2cm]{geometry}

\def\hl{\hspace{7mm}}
\def\hh{\hspace{2mm}}
\def\hs{\hspace{0.5mm}}
\def\hn{\hspace{-10mm}}
\def\DD{\mathrm{D}}
\def\rmd{\mathrm{d}}
\def\rme{\mathrm{e}}

\title{Duality relations between spatial birth-death processes and diffusions in Hilbert space}

\author[$\dag$]{Chris D Greenman\\ \texttt{C.Greenman@uea.ac.uk}}

\affil[$\dag$]{School of computing sciences, University of East Anglia, Norwich Research Park, Norwich, NR4 7TJ.}

\begin{document}

\maketitle

\begin{abstract}
Spatially dependent birth-death processes can be modelled by kinetic models such as the BBGKY hierarchy. Diffusion in infinite dimensional systems can be modelled with Brownian motion in Hilbert space. In this work Doi field theoretic formalism is utilised to establish dualities between these classes of processes. This enables path integral methods to calculate  expectations of duality functions. These are exemplified with models ranging from stochastic cable signalling to jump-diffusion processes.
\end{abstract}

\section{Introduction}

Birth-death processes are concerned with fluctuations in the size of a population of interest, such as growing populations of cells, chemical reactions between molecules, and connectivity of networks, for example. Standard approaches either have no spatial component and are just concerned with population size, or assume the spatially dependent population is fully mixed, with position play no crucial role. However, for many problems of interest, spatial aspects are important. For example, chemical reaction fronts exhibit non-homogeneous spatial behaviour, and incorporating spatial effects into the birth-death interactions is important. 

The theory behind stochastic analysis of fluctuating populations can be traced back to the master equation, originally developed by Kolmogorov \cite{Kolmogoroff1931}. Approaches to birth-death processes underwent crucial development by Kendall \cite{Kendall1948} and Karlin and McGregor \cite{Karlin1957b, Karlin1957a}. A fully stochastic description of systems of spatially interacting particles was collectively first achieved with the BBGKY hierarchy of equations \cite{Bogoliubov1946, Born1949, Kirkwood1946, Kirkwood1947, Yvon1935}. Further development by Doi \cite{Doi1, Doi2} utilized machinery from quantum field theory to examine the dynamics, with path integral approaches for birth-death processes developed by Peliti \cite{Peliti}. System size approaches to mesoscopic analyses were developed by Van Kampen \cite{VanKampen1992}. Comprehensive modern treatments detailing a fuller range of approaches can be found in \cite{Liggett2012, Tauber2014, VanKampen1992}. 

Although the entities in a population of interest can be particulate in nature, such as molecules in a chemical reaction for example, they can also be individuals, queue lengths, cells, network configurations to name a few. However, they shall henceforth be referred to as \emph{particles}. The stochastic dynamics of these populations are characterized by having one or more species of particle, which can increase or decrease in number, due to processes such as birth, death, immigration and emigration, for example. Each particle is also often associated with a set of features of interest, such as position, momentum, age, state, or combinations thereof. These features can be intrinsically discrete, become discrete when lattice approximations to a continuum are considered, or can be continuous. We shall refer all such covariates with a \emph{position}, although this is just a placeholder label that can refer to any feature of interest. The collective set of current positions constitute the state of the system, which are generally Markovian in nature. Note that if the positions are ignored then the state of the system reduces to a particle number count and has the appearance of a classical birth-death process, which spatial processes such as chemical reactions generalize. The techniques to analyse these systems are numerous \cite{Vastola2019a}, but we draw particular attention to Doi-Peliti methods, which are later utilized in this work. 

Doi-Peliti methods allow field theoretic approaches to analyse particle dynamics, and were first introduced by Doi \cite{Doi1, Doi2} with applications to chemical reaction kinetics. A lattice based path integral formulation ws later developed for birth-death processes by Peliti \cite{Peliti}. These papers have seen a plethora of applications; birth-death processes, age structured systems \cite{Greenman3}, neural networks \cite{Buice2007}, algebraic probability \cite{Ohkubo2013b}, knot theory \cite{Rohwer2015}, critical dynamics \cite{Tauber2014}, and phylogeny \cite{Jarvis2005} is a non exhaustive list of examples. Although most path integral approaches are based on the discretization process in \cite{Peliti}, more recent work has seen the development and application of continuous analytic rather than lattice based path integral approaches \cite{Greenman3, Vastola2019b}. Systems with exclusion or partial exclusion properties have also seen the development of fermionic \cite{ Schulz2005, Tauber2014} and parafermionic \cite{Greenman4, Schulz1996} approaches to analysis. 

Diffusion processes, at face value, are seemingly unrelated, and originally arose as models associated with thermodynamics of heat, and also of Brownian motion \cite{Borodin2012}. Models of diffusion can be constructed in any dimension, for example, share prices (one), Brownian motion (two), heat (three), and allelic frequencies of a range of genes ($n \in \mathbb{N}$) in a large population are described by Wright-Fisher diffusion. More recently, techniques have emerged that consider diffusion acting on Hilbert and Banach spaces \cite{Bogachev2011, Bogachev1995, Bogachev2009, Prato2014, Flandoli2016, Gawarecki2010}. Some biological applications can be found in \cite{Prato2014}, including allelic frequencies in a position dependent population, modelled by the infinite dimensional Wright-Fisher diffusions, and the stochastic cable equation that can model neuronal activity \cite{Fleming1975}.  

These seemingly distinct areas of stochastic particle interactions and diffusion processes will be shown to be connected by notions of duality. Duality between stochastic processes provide a connection whereby a certain mean of one process can be related to a mean of another process, usually expressed via expectation of a common duality function. A classic example is the duality between Wright-Fisher diffusion and the Kingsman coalescent process (a death-process related to ancestral convergence under time reversal) \cite{Jansen2014, Mohle1999}.

Approaches to finding dualities between stochastic processes is varied in both method of derivation and application. General background on duality and approaches for interacting particles can be found in \cite{Liggett2012}. A wide class of dualities based on symmetry arguments can be found in \cite{Franceschini2018, Giardina2009, Redig2017}. A comprehensive review of duality in Markov processes was recently provided by Jansen and Kurt \cite{Jansen2014}. A review of the full range of duality in genetic processes can be found in \cite{Carinci2015, Mohle1999}. It is also possible to utilize the discrete Fock spaces of Peliti \cite{Ohkubo2010, Ohkubo2013, Ohkubo2017, Ohkubo2019, Peliti} to derive general dualities between birth-death processes and (mostly) one dimensional diffusions. 

If (discrete) birth-death processes are generalised to spatially dependent systems, it is natural to enquire how (or if) the dualities derived in \cite{Ohkubo2010, Ohkubo2013} generalise. This is the question considered in this work, where we develop approaches to establish duality between spatial birth-death models and infinite dimensional diffusion processes. This utilizes a continuous Doi formalism \cite{Doi1, Doi2, Greenman3, Vastola2019b} rather than the discrete approaches of Peliti \cite{Peliti}. 

The paper is arranged as follows. The next section introduces the Doi formalism of (quantum) field theory used to model both the spatial and diffusion processes. Section 3 introduces infinite dimensional Brownian motion and associated machinery. Section 4 then explains the dualities between these different classes of processes. Section 5 explores path integral expectations of duality functions using a stochastic cable process to exemplify the methods. Section 6 explores some techniques from fermionic field theory to find dualities unavailable by earlier methods. Section 7 considers dualities between pairs of particle systems and between pairs of Hilbert space diffusions. Conclusions complete the work.


\section{Doi formalism}

Here we review the field theoretic machinery of Doi, and utilize it to introduce stochastic particle based and diffusion models. The introduction of machinery shall be relatively brief as more comprehensive introductions can be found in \cite{Doi1, Doi2, Greenman3}.


\subsection{Doi Machinery and Particle Models}

Assume then a population of one species of particle is undergoing stochastic fluctuation. The state of the system at any one point in time is described by a vector of the form ${\bf q}_n$, where $n$ is the current number of particles and component $q_i$ is the position of the $i^\mathrm{th}$ molecule. The term $q_i$ can refer to anything describing the state, such as the age of the $i^\mathrm{th}$ member of a population, or the positions and velocities of a physical molecule (quantum uncertainties being ignored). The term $q_i$, being a component of (emboldened) vector ${\bf q}_n$, is thus an abuse of notation, as $q_i$ can itself be vector valued. However, by defining it as the information associated with one particle (the $i^\mathrm{th}$), the distinction should be clear. We let $\Upsilon$ denote the space of interest, that is, $q_i \in \Upsilon$ and ${\bf q}_n \in \Upsilon^n$. We also suppose that $d$ is the dimension of space $\Upsilon$.

The population will be represented by a ket $\ket{{\bf q}_n}$. This can be defined with the aid of creation and annihilation operators, with  $\ket{{\bf q}_n} = \prod_{i=1}^n\psi_{q_i}^\dag\ket{\varnothing}$, where $\ket{\varnothing}$ denotes the empty state and the operators obey the usual commutation relations
\begin{equation}
[\psi_{q_i},\psi_{q_j}^\dag]=\sum_{\sigma \in S_d}\prod_{k=1}^d\delta((q_i)_k-(q_j)_{\sigma(i)}),
\hl [\psi_{q_i},\psi_{q_j}]=[\psi_{q_i}^\dag,\psi_{q_j}^\dag]=0,
\label{CommRel}
\end{equation}
where we have a sum over all permutations $S_d$ between elements of $q_i$ and $q_j$. The kets $\ket{{\bf q}_n}$ are `pure' states, representing a specific population state. However, the systems we wish to describe are random processes, with a random vector ${\bf q}(t)$ that varies in component values (positions) and in length (population size). To introduce stochastic properties associated with ${\bf q}(t)$ we have a general state of the form
\begin{equation}
\ket{\chi_t} = \sum_{n=0}^\infty\int_{\Upsilon^n} \frac{\rmd {\bf q}_n}{n!} \hs f({\bf q}_n,t)\ket{{\bf q}_n},
\label{ChiDef}	
\end{equation}
where $\frac{f({\bf q}_n,t)}{n!}$ is the probability density for state ${\bf q(t)} = {\bf q}_n$, normalized in the sense that
\begin{equation}
\sum_{n=0}^\infty\int_{\Upsilon^n} \frac{\rmd {\bf q}_n}{n!} \hs f({\bf q}_n,t)=1.
\end{equation}
Note that the integral is over all possible values of ${\bf q}_n \in \Upsilon^n$ meaning we can assume $f$ is symmetric in its arguments, as the anti-symmetric parts will integrate to zero. Then we can interpret $\frac{f({\bf q}_n,t)}{n!}$ as the density associated with a random labelling of the $n$ (indistinguishable) particles. Then, summing over all possible labellings $\sum_{\pi \in S_n}\frac{f(\pi({\bf q}_n),t)}{n!}=f({\bf q}_n,t)$ provides the probability density that the current state ${\bf q}(t)$ is composed of the set of (unordered) particles $\{q_1,q_2,\dots,q_n\}$. This density can be recovered from $\ket{\chi_t}$ via
\begin{equation}
f({\bf q}_m,t) = \braket{{\bf q}_m|\chi_t}.
\label{ChiDefRecov}
\end{equation}

We can then use the creation and annihilation operators to represent interactions of interest. For example, consider the following jump-diffusion processes. If $A_p$ represents a particle at position $p$, we have
\begin{equation}
A_p \hh \substack{R_{pq} \\  \longrightarrow} \hh A_q,\hl
A_p  \hh \substack{D_p \\ \leftrightsquigarrow},
\end{equation}
meaning particles diffuse at position dependent rate $D_p$ and particles jump from positions $p$ to $q$ at rate $R_{pq}$. This is a somewhat trivial model, having fixed particle number and no inter-particle interactions, but will set the scene for duality.

Then, following \cite{Doi1}, the Liouvillian operator describing this process will take the form
\begin{equation}
\mathcal{L} = \int_\Upsilon \rmd p\hs D_p \psi_p^\dag \nabla_p^2 \psi_p+\iint_{\Upsilon^2} \rmd p\hs \rmd q\hs R_{pq}(\psi_q^\dag\psi_p-\psi_p^\dag\psi_p),
\end{equation}
where the dynamics are described formally by the `Heisenberg' evolution equation,
\begin{equation}
\ket{\chi_t} = \rme^{\mathcal{L}t}\ket{\chi_0},
\label{ChiDefDyn}
\end{equation}
for some initial state $\ket{\chi_0}$.

The last technical requirement is the notion of a coherent state. These are needed to calculate averages and construct path integrals, but are also needed to form processes dual to particle systems. Specifically then, for a function $x$ acting on $\Upsilon$ we have coherent state
\begin{equation}
\ket{x} = \exp\left\{\int_{\Upsilon} \rmd p \hs x(p)\psi_p^\dag\right\}\ket{\varnothing}.	
\end{equation}
These act as eigenstates for annihilation and creation operators in the sense that
\begin{equation}
\psi_p\ket{x} = x(p)\ket{x}, \hl \psi_p^\dag\ket{x} = \frac{\delta}{\delta x(p)}\ket{x},
\label{EigRel}
\end{equation}
which can be shown via the commutation relations in Eq. \ref{CommRel}. The later equation contains a functional derivative meant in the sense that $\bra{f} \psi_p^\dag \ket{x} = \frac{\delta}{\delta x(p)}\braket{f|x}$. Note, we shall occasionally make use of the shorthand notation $x_p\equiv x(p)$. The commutation relations can also be used to show that $\braket{x|y}=\rme^{\int_\Upsilon \rmd p\hs xy}$. Thus the states $\ket{x}$ that can be normalised are those that belong to the Hilbert space $L^2(\Upsilon)$.

This is all the machinery that is required for analysing models of interest. For example, the master equation can be derived from the expression $\frac{\partial f({\bf q}_m,t)}{\partial t} = \braket{{\bf q}_m|\mathcal{L}|\chi_t}$. The resulting integral-differential equations often take the form of BBGKY like hierarchies \cite{Bogoliubov1946, Born1949, Kirkwood1946, Kirkwood1947, Yvon1935}. These are generally difficult to solve, and we do not explore these further here (see \cite{Greenman2, Greenman3, Greenman1} for examples).

A slightly simpler problem is to investigate correlation functions for the system. The $m^\mathrm{th}$ order correlation function $X({\bf q}_m)$ represents the probability density for finding $m$
individuals amongst the population with ages given by the set $\{ q_1,q_2,\dots,q_n\}$, and satisfies the dynamic equation $\frac{\partial X({\bf q}_m)}{\partial t}=\braket{1|\prod_{i=1}^m\psi_{q_i}\mathcal{L}|\chi_t}$. This also results in hierarchies of equations, although they tend to be simpler \cite{Greenman2, Greenman3, Greenman1}. Note here that the bra $\bra{1}$ is the coherent state with constant function $1$ (rather than a single particle at position $1$).

One alternative approach to calculate either function $f$ or $X$ is via path integrals. These can be constructed either through spatial discretization techniques first exemplified with the Fock space methods in \cite{Peliti}, or more directly using the Doi framework through continuous techniques \cite{Greenman3}. This gives two approaches for both functions; solving PDEs or calculating path integrals. Although we will not consider these choices to analyse $f$ or $X$ further here, we will later consider both techniques to investigate expectations of duality functions of interest. 


\subsection{Doi Machinery and Diffusion Processes}

Next, a process dual to the one above is constructed. Duality will be established later. Then first we introduce the time dependent state
\begin{equation}
\ket{\Psi_t} = \int \mathcal{D} x \hs P(x,t)\ket{x},
\label{PsiDef} 
\end{equation}
where we have path integration over functions $x$ meant in the sense 
\begin{equation}
\int\mathcal{D}x\hs P(x,t)\ket{x}=\prod_p\int \rmd [x(p)]\hs P(x,t)\rme^{\epsilon\sum_p x(p)\psi_p^\dag}\ket{\phi},
\end{equation}
with positions $p$ taken on a lattice spanning $\Upsilon$ with spacing size $\epsilon$. Eq. \ref{PsiDef} is analogous to Eq. \ref{ChiDef} and  $P(x,t)$ is interpreted as the (infinite dimensional) probability density functional associated with function $x$. Analogously to Eq. \ref{ChiDefRecov}, we can (formally at least) recover this functional via
\begin{equation}
P(x,t) = \int \mathcal{D}y \hs \rme^{-i\int_\Upsilon \rmd p \hs yx}\braket{y|\Psi_t}.
\end{equation} 

Next, analogous to Eq. \ref{ChiDefDyn}, we introduce dynamics with an evolution equation of the form
\begin{equation}
\ket{\Psi_t} = \rme^{\mathcal{L}^\dag t}\ket{\Psi_0}.
\end{equation}
The choice of adjoint operator $\mathcal{L}^\dag$ will become apparent when duality is later considered. 

Then on the one hand, for general bra $\bra{g}$, we find
\begin{equation}
\bra{g}\frac{\partial}{\partial t}\ket{\Psi_t} = \int \mathcal{D} x \hs \frac{\partial P(x,t)}{\partial t}\braket{g|x} 
= \frac{\partial }{\partial t}\int \mathcal{D} x \hs P(x,t) G(x),
\end{equation}
where we have introduced general functional $\braket{g|x}=G(x)$. But we can also construct the following, using the specific adjoint $\mathcal{L}^\dag$ of the operator corresponding to the jump-diffusion model above, to give
\begin{equation}
\bra{g}\frac{\partial}{\partial t}\ket{\Psi_t} = \bra{g}\frac{\partial}{\partial t}\rme^{\mathcal{L}^\dag t}\ket{\Psi_0} = \int \mathcal{D} x \hs P(x,t)\bra{g}\mathcal{L^\dag}\ket{x}.\nonumber \\ 
\end{equation}

Then using the eigen-operator relations in Eq. \ref{EigRel} we find,
\begin{equation}
\frac{\partial}{\partial t} \int \mathcal{D} x \hs P(x,t) G(x)= \int \mathcal{D} x \hs P(x,t)\left\{\int_\Upsilon \rmd p \hs D_p \nabla_p^2(x_p)\frac{\delta}{\delta x_p}\right.
+ \left.\iint_{\Upsilon^2} \rmd p \hs \rmd q \hs R_{pq}  (x_q-x_p)\frac{\delta}{\delta x_p}\right\}G(x).
\label{Peq}
\end{equation}

Thus we have the structure of a Fokker-Planck equation (the form seen for Hilbert spaces \cite{Prato2014, Gawarecki2010, Bogachev2009, Bogachev2011}); $\frac{\partial}{\partial t}\int_H \rmd \mu(x,t) G(x) = \int_H \rmd \mu(x,t) (LG)(x)$, where $\int \rmd \mu(x,t)\equiv \int\mathcal{D}xP(x,t)$ is the measure, $G$ a general functional and we have Kolmogorov operator
\begin{equation}
L = \int_\Upsilon \rmd p \hs D_p \nabla_p^2(x_p)\frac{\delta}{\delta x_p} + \iint_{\Upsilon^2} \rmd p \hs \rmd q \hs R_{pq}  (x_q-x_p)\frac{\delta}{\delta x_p}.
\label{Leg}
\end{equation}

Note that the Kolmogorov operator is obtained in general via the correspondence
\begin{equation}
\bra{g}\mathcal{L}^\dag(\psi_p^\dag,\psi_p)\ket{x} = L\left(\frac{\delta}{\delta x_p},x_p\right)G(x).
\label{KolCorr}
\end{equation}
There are two key things to note with this expression. Firstly, it is assumes that Liouvillian operator $\mathcal{L}^\dag$ is in normal form. That is, the creation operators are left of the annihilation operators. Although all operators discussed will be written in normal form, this can readily be achieved for general forms of operator with the aid of the commutation relations in Eq. \ref{CommRel}. Secondly, the order of operators (following the mapping $\psi_p^\dag \rightarrow \frac{\delta}{\delta x_p}$ and $\psi_q \rightarrow x_q$) are reversed in the Kolmogorov operator. For example $\braket{g|\psi_p^\dag\psi_q|x} = x_q\frac{\delta}{\delta x_p}G(x)$. We are then left with the question of whether the Kolmogorov operator in Eq. \ref{Leg} corresponds to a stochastic process of interest.

For the jump-diffusion process, there exists the probability conservation condition $\bra{1}\mathcal{L} = 0$. This condition is necessary but not sufficient; it does not guarantee positive probabilities, for example. We can apply a similar condition for the function process from Eq. \ref{PsiDef}, where we see that $\bra{\varnothing}\mathcal{L}^\dag = 0$ needs to be satisfied to guarantee $\frac{\partial}{\partial t}\int \mathcal{D}x\hs P(x,t)=0$, which is certainly true for the operator in question. However, this is also not a sufficient condition for a probability process and we need to examine the nature of random function processes in a bit more detail to better understand the operator in question.


\section{Brownian motion in Hilbert space}

Brownian motion in Hilbert space is a well characterized phenomenon and the following brief introduction only covers the salient points. More comprehensive treatments can be found in \cite{Bogachev2011, Bogachev2009, Prato2014, Gawarecki2010, Hairer2009}.

Analogous to Brownian motion in finite dimensional space, a stochastic PDE for Brownian motion in Hilbert space can be written in the form
\begin{equation}
\rmd X_t = (AX_t+Z(X_t))\rmd t + B(X_t)dW_t,
\label{BMIDE}
\end{equation}
where $X_t$ is a stochastic process with values in a separable Hilbert space $H$. The operator $A:H \rightarrow H$ is the infinitesimal generator of a $C_0-$semigroup in $H$. For our purposes we shall largely stick to the case of a diffusion operator $A=D_p\nabla_p^2$. The functions $Z:H\rightarrow H$ and $B:H\rightarrow L_2(H)$ are generally non-linear in nature. Here $W_t$ is a $Q-$Brownian motion taking values in $H$, with
\begin{equation}
W_t = \sum_{i=0}^\infty \lambda_i^{\frac{1}{2}}W_t^{(i)}\xi_i.	
\end{equation}
This is a sum over independent standard one dimensional Weiner processes $W_t^{(i)}$, where $\lambda_i$ and $\xi_i$ are eigenvalues and (orthonormal) eigenfunctions of a trace class operator $Q$. The trace class property means that $\lambda_i \ge 0$, $\sum_{i=0}^\infty \lambda_i < \infty$, and furthermore $\mathrm{Tr}(Q) = \sum_k\braket{Qe_k,e_k}_H$ is finite for any orthonormal basis $e_k$ of $H$ (such as $\xi_k$), with value $\mathrm{Tr}(Q)$ independent of the chosen basis. 

Although the operator $Q=I$ is not trace class, it corresponds at the formal level to the process where $\rmd W_t(p)\rmd W_t(q) = \rmd t \hs\delta(p-q)$ and is known as cylindrical Brownian motion \cite{Hairer2009}.

Note that more generally $W_t$ and $X_t$ can belong to different Hilbert spaces \cite{Gawarecki2010, Prato2014}, and can also be considered in Banach spaces, although we shall not make use of such flexibility here. 

Then the Kolmogorov operator corresponding to the process given in Eq. \ref{BMIDE} is given by \cite{Bogachev2009, Bogachev2011, Gawarecki2010, Prato2014}
\begin{equation}
(LG)(x) = \braket{Ax+Z(x),G_x(x)}_H+\frac{1}{2}\mathrm{Tr}(G_{xx}(x)B(x)QB(x)^*).
\label{IFF}
\end{equation}
Note that the terms $G_x$ and $G_{xx}$ are the functional forms of the Grad and Hessian \cite{Coleman2012}, where we have $\DD G(x)(y)=\braket{G_x(x),y}_H$, $G_{xx}(x) = \DD(G_x)$ and $\DD^2G(x)(y)(z) = \braket{G_{xx}(x)(y),z}_H$, with $\DD$ denoting the Gateaux derivative.

Now if we take generator $A = D_p \nabla_p^2$, the function $Z(x)(p) = \int_\Upsilon \rmd q \hs R_{pq}(x_q-x_p)$, and $B(x) = 0$, the trace term in Eq. \ref{IFF} is zero and 
\begin{eqnarray}
\hn\hn\braket{Ax+Z(x),G_x(x)}_H & = & \DD G(x)(Ax+Z(x)) =  \int_\Upsilon \rmd p \hs \frac{\delta G}{\delta x_p}((Ax)(p)+Z(x)(p)) \nonumber\\
& = & \int_\Upsilon \rmd p \hs D_p \nabla_p^2(x_p)\frac{\delta G}{\delta x_p}
+ \iint_{\Upsilon^2} \rmd p \hs \rmd q \hs R_{pq}  (x_q-x_p)\frac{\delta G}{\delta x_p}.
\label{FFP}
\end{eqnarray}
Thus we have recovered the terms from Eq. \ref{Leg}, and find that $X_t$ is a (deterministic) process of the form
\begin{equation}
\rmd X_t(p) = \left(D_p\nabla^2_pX_t(p) + \int_\Upsilon \rmd q \hs R_{pq}  (X_t(q)-X_t(p))\right)\rmd t + 0\hs\rmd W.
\label{HilbertDeterm}
\end{equation}


\section{Duality}

So far we have introduced two processes, one is a random vector of positions ${\bf q}(t)$ with (stochastic) dynamics described by Liouvillian $\mathcal{L}$, the other is a function $X_t$ with (deterministic) dynamics described by adjoint Liouvillian $\mathcal{L}^\dag$. We now suppose that the processes are initialized with (non-random) vector ${\bf q}(0)={\bf p}_m$ and function $X_0 = z$. We now connect these two processes with the expression $C({\bf p}_m,z;t) = \braket{{\bf p}_m|\rme^{\mathcal{L}^\dag t}|z}$. Although time dependence is present, we shall mostly use the expression $C({\bf p}_m,z)$ unless time is explicitly analysed. One the one hand we find that
\begin{eqnarray}
\label{SimDu}
C({\bf p}_m,z) & = & \braket{{\bf p}_m|\Psi(t)} = \braket{{\bf p}_m|\int \mathcal{D}x\hs P(x,t)|x} \nonumber\\
& = & \int \mathcal{D}x\hs P(x,t)\prod_{i=1}^mx(p_i) = \mathbb{E}_{X}\left(\prod_{i=1}^mX_t(p_i)\right).
\end{eqnarray}

Alternatively, we find that
\begin{equation}
C({\bf p}_m,z) = \braket{\chi(t)|z} = \sum_{n=0}^\infty\int \rmd {\bf q}_n \frac{f({\bf q}_n,t)}{n!}\braket{{\bf q}_n|z} = \mathbb{E}_{\bf q}(z\left({\bf q}(t))\right),
\end{equation}
where we introduce the convention
\begin{equation}
z({\bf q})=\left(\prod_{i=1}^{|{\bf q}|}z(q_i)\right),
\end{equation}
with $|{\bf q}|$ being defined as the length of vector ${\bf q}$.

In summary, we have established a duality between the particle process ${\bf q}(t)$ and the diffusion process $X_t$. Note that the expectation $\mathbb{E}_X$ is an $m^\mathrm{th}$ order correlation function for a set of fixed positions ${\bf p}_m$. Conversely, $\mathbb{E}_{\bf q}$ is a nonlinear expectation, but does offer the freedom to choose $z$.

To calculate $C({\bf p}_m,z)$, we can thus determine either expectation. We will see in subsequent sections that path integrals are one way of doing this. For the example above, a more direct approach is possible. From Eq. \ref{HilbertDeterm} we have a deterministic process and the expectation $\mathbb{E}_X$ is simply the product $\prod_{i=1}^m X_t(q_i)$, where $X_t$ is the solution to the system
\begin{equation}
\left\{
\begin{array}{rcl}
\displaystyle\frac{\partial X_t(p)}{\partial t} & = & D_p\nabla_p^2X_t(p) + \int_\Upsilon \rmd q \hs R(p-q)(X_t(p)-X_t(q)),\\
X_0(p) & = & z.
\end{array}
\right.
\end{equation}

If we take the homogeneous model, where $D_p=D$ is constant and $R_{pq}=R(p-q)$ just depends on separation, such that the total jump rate from any given position $R_{tot} = \int \rmd r \hs R(r) < \infty$ is finite, this equation is straightforwardly solved via Fourier transform techniques, where we find $X_t(p) = \mathcal{F}_p^{-1}(\rme^{(R(q)-R_{tot}-q^2D)t}\mathcal{F}_q(z(p)))$. Then the duality condition implies expectation $\mathbb{E}_{\bf q}$ is just a product over these functions. This is to be expected; firstly, jump-diffusion is non-interacting, so independent across the $m$ particles involved in the process, and secondly, the expected position will be dictated by the bias in the jumping function $R_{pq}$, and the diffusion of weight function $z$.


\section{Stochastic Cable Equation and Path Integral Methods}

For the next example, we start with a stochastic PDE in Hilbert space and instead search for a dual system of particle interactions. Path integral techniques and dynamic approaches will then be used to calculate the duality expectations, where we will find a Feynman-Kac term is needed.


\subsection{Cable Equation and Dual Process}

Consider then the following cable equation, which is a stochastic Nagumo equation used to model neuronal excitations \cite{Prato2014, Mckean1970},

\begin{equation}
\rmd X_t(p) = (\nabla_p^2 X_t - X_t)\rmd t + \rmd W_t,
\label{CabSpde}
\end{equation}
where $\rmd W_t$ represents $Q$-Brownian motion. This has a Kolmogorov operator of the form (compare with Eq. \ref{IFF})
\begin{eqnarray}
(LG)(x) & = & \braket{\nabla_p^2 x - x,G_x(x)}_H + \frac{1}{2}\mathrm{Tr}(G_{xx}(x)Q)\nonumber\\
& = & \DD G(x)(\nabla_p^2(x)) - \DD G(x)(x) + \frac{1}{2}\sum_k\lambda_k\DD^2 G(x)(\xi_k)(\xi_k),
\label{FF2}
\end{eqnarray} 
where $\lambda_k$ and $\xi_k$ are eigenvalues and eigenfunctions of trace-class operator $Q$. 

Next consider the operator defined by
\begin{equation}
\mathcal{L}^\dag = \int_\Upsilon \rmd p \hs \psi_p^\dag \nabla_p^2\psi_p
- \int_\Upsilon \rmd p \hs \psi_p^\dag\psi_p 
+ \iint_{\Upsilon^2} \rmd p \hs \rmd q \hs R_{pq} \psi_p^\dag\psi_q^\dag.
\end{equation}
Then we find from Eq. \ref{KolCorr} a Kolmogorov operator of the form
\begin{equation}
(LG)(x) = \int_\Upsilon \rmd p \hs \frac{\delta G}{\delta x_p}(\nabla^2_p(x)-x_p)
+\iint_{\Upsilon^2} \rmd p \hs \rmd q \hs \frac{\delta^2G}{\delta x_p \delta x_q}R_{pq}.
\end{equation}

Now, if we define the symmetric function $R_{pq} =R(p,q)=\frac{1}{2}\sum_k\lambda_k\xi_k(p)\xi_k(q)$, we obtain Eq. \ref{FF2}. 

Conversely, we can write the Liouvillian operator as $\mathcal{L} = \mathcal{L'}+V$, with operators
\begin{eqnarray}
\mathcal{L'} = \int_\Upsilon \rmd p \hs \psi_p^\dag \nabla^2_p\psi_p 
 + \iint_{\Upsilon^2} \rmd p \hs \rmd q \hs R_{pq} \left(\psi_p\psi_q-\psi_p^\dag\psi_q^\dag\psi_p\psi_q\right), \nonumber\\
V= - \int_\Upsilon \rmd p \psi_p^\dag\psi_p+ \iint_{\Upsilon^2} \rmd p \hs \rmd q \hs R_{pq} \psi_p^\dag\psi_q^\dag\psi_p\psi_q.
\label{Eldash}
\end{eqnarray}
Then $\mathcal{L'}$ is the evolution operator of the diffusion-annihilation particle process
\begin{equation}
A_p + A_q \hh \substack{R_{pq} \\  \longrightarrow} \hh \phi,\hl
A_p  \hh \substack{D_p \\ \leftrightsquigarrow},
\end{equation}
which represent particles diffusing at rate $D_p$ and pairwise annihilating at rate $R_{pq}$, where $p$ and $q$ represent the positions of the two annihilating particles. Note that the antisymmetric part of $R_{pq}$ integrates to zero in Eq. \ref{Eldash} and we can assume $R_{pq}$ is symmetric in its two arguments, which is also a natural assumption for the process. 

Two processes have now been constructed; $X_t$ is the cable process, which we initialize with some function $z$, and ${\bf q}(t)$ is the diffusion-annihilation process, starting from some vector ${\bf p}_m$. The two processes do not quite have adjoint evolution operators, so a little work is needed to construct duality.


\subsection{Duality and Feynman-Kac Form}

Duality is again constructed from $C({\bf p}_m,z) = \braket{{\bf p}_m|\rme^{\mathcal{L}^\dag t}|z}$, where on the one hand we have, $C({\bf p}_m,z)=\mathbb{E}_X\left(X_t({\bf p}_m)\right)$ in exactly the same manner as Eq. \ref{SimDu}.

To gain a dual expectation requires the derivation of a Feynman-Kac term. This is done via path integration, which requires the following resolution of the identity $I$ (verifiable by showing $I\ket{{\bf q}_n}=\ket{{\bf q}_n}$ via the commutation relations of Eq. \ref{CommRel}),
\begin{equation}
I = \sum_k\int_{\Upsilon^k}\frac{\rmd {\bf r}_k}{k!} \hs \ket{{\bf r}_k}\bra{{\bf r}_k} = \int_{\hat\Upsilon} \rmd \hat{{\bf r}}\ket{\hat{\bf r}}\bra{\hat{\bf r}}.
\end{equation}
The right hand side is notation to represent integration of vectors $\hat{\bf r}$ over the space $\hat\Upsilon = \cup_{k=0}^\infty\Upsilon^k$, so both coordinates and lengths of the vectors vary, introduced to simplify notation below. The resolutions of identity can then be used to interlace $N$ time slices of width $\epsilon$, where we find
\begin{eqnarray}
\label{DPIR}
C({\bf p}_m,z) & = & \braket{z|\rme^{(\mathcal{L}'+V) t}|{\bf p}_m}
= \braket{z|\prod_{k=1}^N\rme^{(\mathcal{L}'+V) \epsilon}|{\bf p}_m}\nonumber\\
& = & \prod_{k=0}^N \int \rmd \hat{\bf r}_k \braket{z|\hat{\bf r}_N}\prod_{\ell=1}^N\braket{\hat{\bf r}_\ell|\rme^{\mathcal{L}'\epsilon+V\epsilon}|\hat{\bf r}_{\ell-1}}\braket{\hat{\bf r}_0|{\bf p}_m} \nonumber\\
& = & \prod_{k=0}^N \int \rmd \hat{\bf r}_k \rme^{\epsilon\sum_{l=1}^NV(\hat{\bf r}_l)} \braket{z|\hat{\bf r}_N}\prod_{\ell=1}^N\braket{\hat{\bf r}_\ell|\rme^{\mathcal{L}'\epsilon}|\hat{\bf r}_{\ell-1}}\braket{\hat{\bf r}_0|{\bf p}_m}.
\end{eqnarray}
Some observations help interpret this expression. Firstly, we have function $V(\hat{\bf r}) = -|\hat{\bf r}|+\sum_{i \ne j}R_{\hat{r}_i\hat{r}_j}$, which is found from the action of operator $V$ on pure state kets $\ket{\hat{\bf r}}$. The distinction between operator $V$ and function $V(\hat{\bf r})$ will hopefully be clear from the context it is found. Secondly, the term $\braket{\hat{\bf r}_0|{\bf q}_m}$ reduces to delta functions via Eq. \ref{CommRel}, so that integration over $\hat{\bf r}_0$ forces the initial condition $\hat{\bf r}_0={\bf q}_m$. Thirdly, also via Eq. \ref{CommRel}, we get the term $\braket{x|\hat{\bf r}_N} = x(\hat{\bf r}_N)$. Fourthly, we note from Eq. \ref{ChiDefRecov} that $\braket{\hat{\bf r}_\ell|\rme^{\mathcal{L}'\epsilon}|\hat{\bf r}_{\ell-1}}$ is simply the probability density for transition from state $\hat{\bf r}_{\ell-1}$ to $\hat{\bf r}_{\ell}$. Then $\prod_{\ell=1}^N \braket{\hat{\bf r}_\ell|\rme^{\mathcal{L}'\epsilon}|\hat{\bf r}_{\ell-1}} $ is just the joint probability density of the path $\hat{\bf r}_{0}\rightarrow \hat{\bf r}_{1}\rightarrow\dots\rightarrow \hat{\bf r}_{N}$ conditional on start vector $\hat{\bf r}_{0} = {\bf p}_m$. Finally, in the continuum limit we formally write $P[\hat{\bf r}]$ for the density associated with the path $\hat{\bf r}(s), s \in [0,t]$ and obtain path integral
\begin{eqnarray}
\label{PINC}
C({\bf p}_m,z) & = & \int_{\hat{\bf r}(0) = {\bf p}_m} \mathcal{D} \hat{\bf r}\hs P[\hat{\bf r}]z(\hat{\bf r}(t))\exp\left\{\int_0^t\rmd s \hs V(\hat{\bf r}(s))\right\}\nonumber\\
& = & \mathbb{E}_{\bf q}\left(z({\bf q}(t))\exp\left\{{\int_0^t}\rmd s \hs V({\bf q}(s))\right\}\right),
\end{eqnarray}
Note that this construction only works if pure states $\ket{\hat{\bf r}}$ are eigenstates of operator $V$. A sufficient condition is to require that $V$ can be written as a function (or functional) of $\psi_q^\dag\psi_q$, which is true for this example. 

Thus we have duality $C({\bf p}_m,z) = \mathbb{E}_X(X_t({\bf p}_m)) = \mathbb{E}_{\bf q}\left(z({\bf q}(t))\exp\left\{{\int_0^t}\rmd s \hs V({\bf q}(s))\right\}\right)$, where process $X_t$ is initialized with $X_0=z$ and process ${\bf q}(t)$ with ${\bf q}(t)={\bf p}_m$. The expectation is complicated by the Feynman-Kac functional $\int_0^t\rmd s\hs V({\bf q}(s))$ \cite{Ohkubo2013, Borodin2012} and although we have a form of duality, there is no duality function in the classical sense.


\subsection{Coherent State Path Integration}

Note that the path integral in the previous section is over paths taken by a vector $\hat{\bf r}(s)$ which varies in value and length. This is not the usual form of path integrals used for calculations in the Doi framework, which tend to be based on coherent states (or equivalently through the Bargmann-Fock space approach used by Peliti \cite{Peliti, Itzykson2012}), which we now make use of. We next show that $C({\bf p}_m,z)$ can be calculated exactly with the aid of a coherent state path integral \cite{Greenman3}. This requires an alternative resolution of identity $I = \iint \mathcal{D}u\hs\mathcal{D}v\hs\rme^{-i\int_\Upsilon \rmd q \hs uv}\ket{iv}\bra{u}$ to demarcate $N$ time intervals of width $\epsilon$ and construct the following
\begin{eqnarray}
\hn\hn C({\bf p}_m,z) & = & \braket{{\bf p}_m|\rme^{\mathcal{L}^\dag t}|z}
=\braket{{\bf p}_m|\prod_{k=1}^N\rme^{\mathcal{L}^\dag \epsilon}|z}\nonumber\\
& = & \prod_{k=0}^N\iint \mathcal{D}u_k \hs \mathcal{D}v_k\rme^{-i\sum_{k=0}^N\int_\Upsilon \rmd q \hs u_iv_i}\braket{{\bf p}_m|iv_N}
\prod_{k=1}^N\braket{u_k|\rme^{\mathcal{L}^\dag \epsilon}|iv_{k-1}}
\braket{u_0|z}\nonumber\\
& = & \iint \mathcal{D}u \hs \mathcal{D}v \hs \prod_{j=1}^m [iv(p_j,t)]\exp\left\{ i\int_0^t \rmd s \hs\int_\Upsilon \rmd q\hs\hs v\left(\nabla_q^2 u - u + \frac{\partial u}{\partial s}\right) \right.\nonumber\\
&& \hl \left.-i\int_\Upsilon \rmd q\hs u(q,t)v(q,t)+\int_0^t\rmd s\hs\iint_{\Upsilon^2}\rmd q\hs \rmd q'\hs R_{qq'}u(q,s)u(q',s) \right\}.\nonumber\\
&& \hl\hl\hl\exp\left\{\int_\Upsilon \rmd q\hs u(q,0)z(q)\right\}.
\label{BigC}
\end{eqnarray}
The last step makes use of eigenfunction properties of coherent states given in Eq. \ref{EigRel} to form the final path integral. For example, if $\mathcal{L}^\dag$ is in normal form, $\braket{u_k|\mathcal{L}^\dag (\psi_q^\dag,\psi_q)|iv_{k-1}}=\braket{u_k|iv_{k-1}}\mathcal{L}^\dag(u_k(q),iv_{k-1}(q))=\rme^{i\int_\Upsilon \rmd q\hs u_kv_{k-1}}\mathcal{L}^\dag(u_k(q),iv_{k-1}(q))$. Note that variables $u$ and $v$ in the final path integral are integrated over both time $t$ and space $q$.

\begin{figure}[t]
\centering
\includegraphics[width=14cm]{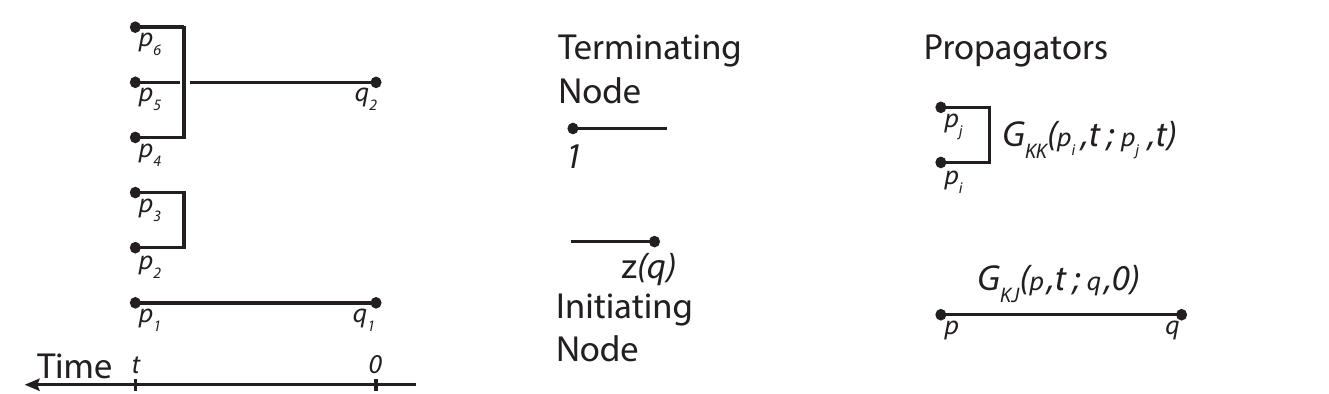}
\caption{Feynman diagram information for duality function $C({\bf p}_m,z)$ of the cable equation. A sample diagram, initiating and terminating node types, and associated propagators are provided.}
\label{Feynman}
\end{figure}

This can now be treated perturbatively. The terms $\prod_{j=1}^m [iv(p_j,t)]$ will correspond to $m$ termination nodes in a pertubative expansion Feynman diagram (see Fig. \ref{Feynman}), and expanding $\rme^{\int_\Upsilon \rmd q\hs u(q,0)z(q)}$ will result in any number of initiating nodes with coefficient $z(q)$. There will be no internal nodes as the remaining part of the path integral can be calculated directly and absorbed into propagators by using a generating functional of the form
\begin{eqnarray}
\hn Z(J,K) & = & \iint \mathcal{D}u \hs \mathcal{D}v \hs \exp\left\{ \int_0^t \rmd s \hs\int_\Upsilon \rmd q \hs iv\left(\nabla_q^2 u - u + \frac{\partial u}{\partial s}\right)\right. \nonumber\\
&& -i\int_\Upsilon \rmd q \hs u(q,t)v(q,t)
+\int_0^t\rmd s \hs\iint_{\Upsilon^2} \rmd q \hs \rmd q' \hs R_{qq'}u(q,s)u(q',s)\nonumber\\
&& \left. +\int_0^t \rmd s\hs \int_\Upsilon \rmd q \hs (uJ+ ivK)\right\}.
\end{eqnarray}
Now, integrating over the $v$ variable gives delta functionals that restrict $u$ to a form obeying
\begin{equation}
\left\{	
\begin{array}{rcl}
-\displaystyle\frac{\partial u}{\partial t} & = & \nabla_q^2 u - u + K,\nonumber\\
u(t) & = & 0.
\end{array}
\right.
\end{equation}
Thus we have a reverse time heat equation which can be solved with standard techniques. Note that the space $\Upsilon$ has not yet been specified. Although this equation can be solved in $\mathbb{R}^n$, results in $\mathbb{R}$ are largely similar, where we find
\begin{equation}
u(q,t) = \int_\tau^t \rmd s \hs\int_\mathbb{R} \rmd r \hs \frac{1}{\sqrt{4\pi(s-\tau)}}\exp\left\{ -\frac{(q-r)^2}{4(s-\tau)}\right\}K(r,s)\rme^{-(s-\tau)},
\end{equation}
and we find a generating functional of the form
\begin{equation}
Z(J,K) = \exp\left\{ \int_0^t\rmd s\hs \iint_{\mathbb{R}^2} \rmd q \hs \rmd q' \hs R_{qq'}u(q,s)u(q',s)
+\int_0^t\rmd s\hs\int_\mathbb{R} \rmd q \hs u(q,s)J(q,s) \right\}.
\end{equation}

Thus we find that there are two non-zero propagators that relate to initiating or terminating nodes in the corresponding Feynman diagrams. Specifically, 
\begin{eqnarray}
\hn G_{KK}(p_i,t;p_j,t) & = & \iint \mathcal{D}u\mathcal{D}v\hs iv(p_i,t)\hs iv(p_j,t)\rme^S = 
\left.\frac{\delta^2 Z(J,K)}{\delta K(p_i,t)\delta K(p_j,t)}\right|_{J\equiv K\equiv 0}\nonumber\\
& = & 2 \int_0^t \rmd s\hs \iint_{\mathbb{R}^2}\rmd r\hs \rmd r'\hs\frac{R_{rr'}}{4\pi s}
\exp\left\{\frac{-(p_i-r)^2-(p_j-r')^2}{4s}\right\}\rme^{-2s}\nonumber\\
\hn G_{KJ}(p,t;q,0) & = & \iint \mathcal{D}u\mathcal{D}v\hs iv(p,t)\hs u(q,0)\rme^S =
\left.\frac{\delta^2 Z(J,K)}{\delta K(p,t)\delta J(q,0)}\right|_{J\equiv K\equiv 0}\nonumber\\
& = & \frac{1}{\sqrt{4\pi t}}\exp\left\{\frac{-(p-q)^2}{4t}\right\}\rme^{-t},
\end{eqnarray}
where $S$ is the action, that is, the first exponent of the path integral in Eq. \ref{BigC}. Note that the first propagator is just a time weighted diffusion of the function $R_{p_ip_j}\equiv R(p_i,p_j)$ in both coordinates, and corresponds to the arced edges in Fig. \ref{Feynman}. An arc connecting $p_i$ to $p_j$ thus has an associated factor $2\int_0^t\rmd s\hs \rme^{-2s}(\Phi_s^1\Phi_s^2R)(p_i,p_j)$, where $\Phi_s^kR$ represents the diffusion operator acting on coordinate $k$ of function $R$. 

The second propagator is a time weighted heat kernel, and are associated with initiating nodes which have a factor of the form $z(q)$ where position $q$ is integrated over. Thus a horizontal line in Fig. \ref{Feynman} terminating in $p_i$ has a contribution of the form
\begin{equation}
\int_{\mathbb{R}} \rmd q\hs z(q) G_{KJ}(p_i,t;q,0) = \int_\mathbb{R} \rmd q \hs \frac{z(q)}{\sqrt{4\pi t}}\exp\left\{\frac{-(p_i-q)^2}{4t}\right\}\rme^{-t} = \rme^{-t}(\Phi_t z)(p_i),
\end{equation} 
where $\Phi_t z$ represents the diffusion operator acting on function $z$.

Finally, summing over all possible diagrams, we find
\begin{equation}
C({\bf p}_m,z) = \sum_{P\subset \{1,2,\dots,m\}}
\prod_{\{i,j\} \in P}2\int_0^t\rmd s\hs \rme^{-2s}(\Phi_s^1\Phi_s^2R)(p_i,p_j)
\prod_{k \in P^c}\rme^{-t}(\Phi_t z)(p_k),
\end{equation}
where the subsets $P$ are those containing non-intersecting pairs of distinct elements from the set $\{1,2,\dots,m\}$. 

Note that the order one correlation function $C(p,z)=\rme^{-t}(\Phi_t z)(p)$ is just a simple diffusion of $z$, as can be seen by taking the expectation of Eq. \ref{CabSpde}.


\subsection{Dynamic Duality Equations}

The previous section saw a path integral approach to the calculation of $C({\bf p}_m,z;t) = \braket{{\bf p}_m|\rme^{\mathcal{L}^\dag t}|z}$. However, we can also derive two dynamic equations for $C({\bf p}_m,z;t)$. Then differentiating, we find
\begin{equation}
\frac{\partial C}{\partial t} =  \braket{{\bf p}_m|\rme^{\mathcal{L}^\dag t}\mathcal{L}^\dag|z}
= \braket{z|\rme^{\mathcal{L} t}\mathcal{L}|{\bf p}_m}.
\end{equation}

Now on the one hand we can let $\mathcal{L}$ act on pure state $\ket{{\bf p}_m}$ to give (via the commutation relations in Eq. \ref{CommRel}),
\begin{equation}
\frac{\partial C}{\partial t} = \nabla_{{\bf p}_m}^2 C - mC + \sum_{i \ne j}R_{p_ip_j}C({\bf p}_m^{-(i,j)},z;t),
\label{df1}
\end{equation}
where ${\bf p}_m^{-(i,j)}$ is the vector ${\bf p}_m$ with components $i$ and $j$ removed. Thus we have a BBGKY like hierarchy of finite dimensional heat equations. Note that unlike many BBGKY hierarchies, the equation for $C({\bf p}_m,z)$ implicates the function $C({\bf p}_m^{-(i,j)},z)$ which depends upon a smaller vector ${\bf p}_m^{-(i,j)}$, meaning these equations could be treated recursively.

Alternatively, we can let $\mathcal{L}^\dag$ act on coherent state $\ket{z}$ to give (via the eigenstate relations in Eq. \ref{EigRel}),
\begin{equation}
\frac{\partial C}{\partial t} = \int_\Upsilon \rmd q \hs\frac{\delta C}{\delta z_q}\left(\nabla_q^2(z)-z \right)
+\iint_{\Upsilon^2} \rmd q \hs \rmd q' \hs\frac{\delta^2C}{\delta z_q \delta z_{q'}}.
\label{df2}
\end{equation}
Thus we have essentially recovered the cable equation of Eq. \ref{FF2}, albeit with different initial conditions; Eq.s \ref{df1} and \ref{df2} both have initial condition $C({\bf p}_m,z;0) = z({\bf p}_m)$.

So to analyse the duality expectations such as $C({\bf p}_m,z;t)$, one can either treat them dynamically to extract differential equations, or try a path integral approach.

We next consider a situation where a Feynman-Kac correction is not sufficient to extract a duality relationship, and Fermionic approaches are needed to deal with unwanted negative signs in evolution operators that seem to rule out a dual process.


\section{Simple Decay with Fermionic Duality Mechanism}

Consider the following stochastic PDE for Brownian motion in Hilbert space $H$
\begin{equation}
\rmd X_t(p) = -\gamma_p X_t(p)^2 \rmd t + B(X_t)\rmd W_t(p),
\end{equation}
where we have Q-Brownian motion $\rmd W_t$, a positive, position dependent decay function $\gamma_p$, and initial function $X_0=z$. For this example we have non-trivial noise, with linear operator $B(X_t)$ defined by its action on the $Q$ eigenfunctions, where we have $B(X_t)(\xi_k)(p) = \sum_m\beta_{km}(X_t)\xi_m(p)$, where we will later specify functionals $\beta_{km}(X_t)$.

This example has been chosen because firstly the negative term will be shown to preclude any obvious duality with a particle process under the framework of previous sections, and secondly the noise term is not a relatively simple point-wise operator of the form $(B(X_t)\xi_k)(p)=b(X_t(p))\xi_k(p)$. That is, $B(X_t)(\xi_k)(p)$ depends globally on the function $X_t$, not just on the value $X_t(p)$.

Then this gives us a Kolmogorov operator of the form:
\begin{eqnarray}
\label{CFP1}
\hn\hn(LG)(x) & = & \braket{-\gamma_p x^2,G_x(x)} + \frac{1}{2}\mathrm{Tr}(G_{xx}B(x)QB(x)^*)\\
& = & -\int_\Upsilon \rmd p \frac{\delta G}{\delta x_p}\gamma_px_p^2 + \frac{1}{2}\iint_{\Upsilon^2} \rmd p \hs \rmd q \frac{\delta^2 G}{\delta x_p \delta x_q}\sum_{k,m,n}\lambda_k\beta_{km}(x)\beta_{kn}(x)\xi_m(p)\xi_n(q).\nonumber
\end{eqnarray}
To connect this operator via Eq. \ref{KolCorr} to an adjoint Liouvillian operator $\mathcal{L}^\dag$, we introduce
\begin{equation}
\mathcal{L}^\dag = -\int_\Upsilon \rmd p \hs\gamma_p \psi_p^\dag\psi_p^2
+\iiint_{\Upsilon^3} \rmd p \hs \rmd q \hs \rmd r \hs R_{pqr}\psi_p^\dag\psi_q^\dag\psi_r.
\label{firstdag}
\end{equation}
Now, much like the previous section, we can compensate the second term to get a dual particle process with Liouvillian $\mathcal{L}'=\mathcal{L}-V$ for a suitable term $V$. However, the negative sign of the first term in Eq. \ref{firstdag} means a compensating term of the form $\int \rmd p \hs \gamma_p\psi_p^\dag\psi_p$ has a positive sign. This means that although the requisite equation $\bra{1}\mathcal{L}'=\bra{1}\left(-\int_\Upsilon \rmd p \hs \gamma_p ((\psi_p^\dag)^2\psi_p -\psi_p^\dag\psi_p) +\dots\right)=0$ is satisfied, and ensures total probability conservation for the dual particle process, the corresponding master equation has the wrong sign and the associated `probabilities' are not necessarily positive.

Instead, following \cite{Ohkubo2013}, we introduce an operator to flip the sign. This stems from algebraic probability arguments in \cite{Ohkubo2013}, however, we will see this can also be framed in terms of a Fermionic Doi algebra \cite{Greenman4}. 

Specifically, we introduce the self adjoint operator $b=a+a^\dag$ where $a$ and its adjoint $a^\dag$ are standard Pauli operators acting on the two dimensional space $\{\ket{0},\ket{1}\}$, satisfying standard anticommutativity relations $\{a,a^\dag\} = 1$ and $\{a,a\} = \{a^\dag,a^\dag\} = 0$. We also introduce orthonormal states $\ket{\pm}=\frac{1}{2}(\ket{0} \pm \ket{1})$, where we note that $b\ket{\pm}=\pm\ket{\pm}$.

Next we replace the Liouvillian in Eq. \ref{firstdag} with
\begin{equation}
\mathcal{L}^\dag = \int_\Upsilon \rmd p \hs \gamma_p \hs b\psi_p^\dag\psi_p^2 + \iiint_{\Upsilon^3} \rmd p \hs \rmd q \hs \rmd r \hs R_{pqr}\psi_p^\dag\psi_q^\dag\psi_r,
\label{SecondDag}
\end{equation}
and introduce dynamics $\ket{\Psi_t}=\rme^{\mathcal{L}^\dag t}\ket{\Psi_0}$ for some initial state $\ket{\Psi_0}$, where, analogous to Eq. \ref{PsiDef}, the states have a representation of the form
\begin{equation}
\ket{\Psi_t} = \int \mathcal{D}x P(x,t)\ket{x,-}.
\end{equation}
Note that strictly speaking the state $\ket{x,-}$ is shorthand for the tensor product $\ket{x}\otimes\ket{-} \in H\otimes \{\pm \}$, and the operators $\psi_p$ and $b$ commute because they act independently on $H$ and $\{\pm\}$. Shorthand rather than formal notation is used throughout.

Now, direct calculation shows us that for generic bra $\bra{g}$ and functional $G(x)=\braket{g|x,-}$,
\begin{eqnarray}
\hn\hn&&\hn\hn\bra{g}\frac{\partial}{\partial t} \int \mathcal{D}x\hs P(x,t)\ket{x,-} =
\frac{\partial}{\partial t} \int \mathcal{D}x\hs P(x,t)G(x) = 
\int \mathcal{D}x \hs P(x,t)\bra{g}\mathcal{L}^\dag\ket{x,-} = \nonumber\\   
&& \hn=  \int \mathcal{D}x\hs P(x,t)\left\{-\int_\Upsilon \rmd p \hs \frac{\delta G}{\delta x_p}\gamma_p x_p
+\iint_{\Upsilon^2} \rmd p \hs \rmd q \hs \frac{\delta^2 G}{\delta x_p \delta x_q}\int_\Upsilon \rmd r \hs R_{pqr}x_r\right\}.
\end{eqnarray}

Thus we obtain the Fokker-Planck equation corresponding to the Kolmogorov operator given in Eq. \ref{CFP1}, provided we have the match
\begin{equation}
\int_\Upsilon \rmd r \hs R_{pqr}x_r = \frac{1}{2}\sum_k\lambda_k(B(x)\xi_k)(p)(B(x)\xi_k)(q).
\label{Match1}
\end{equation}
This offers a wide choice for $R_{pqr}$. Note that the right hand side dictates that $R_{pqr}$ needs to be symmetric in $p$ and $q$ (which will be seen below to also be a natural assumption for a dual particle interaction). To specify $R_{pqr}$ in terms of operator $B$ the orthonormal basis $\xi_k$ can be used. Firstly, write $\int_\Upsilon \rmd r \hs R_{pqr}x_r = \sum_{m,n}\alpha_{mn}(x)\xi_m(p)\xi_n(q)$ for coefficients $\alpha_{mn}(x)$, which are also functionals in $x$. Secondly, we have the earlier assumption that $(B(x)\xi_k)(p) = \sum_m \beta_{km}(x)\xi_m(p)$ for functional coefficients $\beta_{km}(x)$. Then using orthonormality, Eq. \ref{Match1} reduces to equivalent condition
\begin{equation}
\alpha_{mn}(x) = \frac{1}{2}\sum_k\lambda_k\beta_{km}(x)\beta_{kn}(x),
\end{equation}
and the action of $R_{pqr}$ on $x_r$ can be specified in terms of the action of $B(x)$ on the eigenfunctions.

Next for a dual process, we consider the following particle model. We again have particles $A_p$ at position $p$, but now also an overall system-wide state sign $\kappa \in \{\pm\}$. Then we have process
\begin{eqnarray}
A_p \hs\longrightarrow\hs A_p + A_p, \hl \kappa \hs\longrightarrow \hs -\kappa, \hl(\mathrm{rate }\hh\gamma_p)\nonumber\\
A_p + A_q\hs\longrightarrow\hs A_r, \hl \kappa \hs\longrightarrow \hs \kappa. \hl\hh(\mathrm{rate }\hh R_{pqr})
\end{eqnarray}
Thus we have localised particle fission with sign flip, and a non-local pairwise amalgamation process that preserves the system sign. The state of the system $({\bf q},\kappa)$ is next represented by a ket $\ket{{\bf q},\kappa}$. This has a corresponding Liouvillian $\mathcal{L}'=\mathcal{L}-V$ of the form
\begin{equation}
\mathcal{L}' = \int_\Upsilon \rmd p \hs \gamma_p (b(\psi_p^\dag)^2\psi_p-\psi_p^\dag\psi_p) + \iiint_{\Upsilon^3} \rmd p \hs \rmd q \hs \rmd r \hs R_{pqr}(\psi_r^\dag\psi_p\psi_q-\psi_p^\dag\psi_q^\dag\psi_p\psi_q),
\end{equation} 
where $V = \int_\Upsilon \rmd p \hs \gamma_p \psi_p^\dag\psi_p 
+\iiint_{\Upsilon^3} \rmd p \hs \rmd q \hs \rmd r \hs R_{pqr}\psi_p^\dag\psi_q^\dag\psi_p\psi_q$. This results in an evolution equation $\ket{\chi_t}=\rme^{\mathcal{L}'t}\ket{\chi_0}$ for a designated initial state $\ket{\chi_0}=\ket{{\bf p}_m,-}$.

Then analogous to \cite{Ohkubo2013}, we obtain the duality
\begin{equation}
C({\bf p}_m,z) = \braket{{\bf p}_m,-|\rme^{\mathcal{L}^\dag t}|z,-} = \mathbb{E}_X\left(X_t({\bf p}_m)\right) = \mathbb{E}_{({\bf q},\kappa)}\left( z({\bf q}(t))\mathbb{I}_{\{\kappa(t)\equiv -\}}\rme^{\int_0^t \rmd s\hs V({\bf q}(s))}\right).
\label{FermDualEx}
\end{equation}
The last term in this expression is derived in much the same way as Eq. \ref{DPIR}, except that the resolution of identity in this case is $I = \sum_{\kappa \in \{\pm\}}\int_{\hat\Upsilon}\rmd \hat{\bf r}\ket{\hat{\bf r},\kappa}\bra{\hat{\bf r},\kappa}$. The main difference in the subsequent derivation is that the term $\braket{x|\hat{\bf r}_N}$ in Eq. \ref{DPIR} becomes $\braket{z,-|\hat{\bf r}_N,\kappa_N} = \braket{z|\hat{\bf r}_N}\mathbb{I}_{\{\kappa_N \equiv -\}}$ resulting in the form above.

To calculate the expectations one can again use path integrals. We can firstly use the Liouvillian operator in Eq. \ref{SecondDag}. This involves the operator $b=a+a^\dag$, which would implicate a hybrid path integral containing grassmannians \cite{Greenman4} and bosonic integrals \cite{Greenman3}. Alternatively, we can use the original operator in Eq. \ref{firstdag} to construct a path integral. Because both diffusion processes have the same Fokker-Planck equation and so distribution for the $X_t$ process, the expectations of Eq. \ref{FermDualEx} will be the same for both processes, so using the path integral without grassmannians will be simpler.


\section{Particle-Particle and Diffusion-Diffusion Dualities}

Finally, we consider two cases; particle models dual with particle models and Hilbert space diffusions dual with Hilbert space diffusions. 


\subsection{Particle Models}

Consider a simple particle model for budding-birth-death, where the parent particle survives birth of a daughter particle,
\begin{equation}
A_p \hh \substack{\mu_p \\  \longrightarrow} \hh \varnothing,\hl
A_p \hh \substack{\beta_{pq} \\ \longrightarrow} \hh A_p + A_q.
\label{ParPar}
\end{equation}
This has an evolution operator of the form
\begin{equation}
\mathcal{L}' = \int_\Upsilon \rmd p \hs \mu_p(\psi_p-\psi_p^\dag\psi_p)
 +\iint_{\Upsilon^2} \rmd p \hs\rmd q \hs \beta_{pq}(\psi_p^\dag\psi_q^\dag\psi_p-\psi_p^\dag\psi_p),
\end{equation}
resulting in state evolution equation $\ket{\chi_t} = \rme^{\mathcal{L}'t}\ket{\chi_0}$, where we have initial (pure) state $\ket{\chi_0} = \ket{{\bf p}_m}$ and current state with probabilistic interpretation $\ket{\chi_t} = \sum_k\int_{\Upsilon^k} \frac{\rmd {\bf r}_k}{k!} \hs f_{\mathrm{BBD}}({\bf r}_k)\ket{{\bf r}_k} = \int_{\hat\Upsilon} \rmd \hat{\bf r} \hs f_{\mathrm{BBD}}(\hat{\bf r})\ket{\hat{\bf r}}$. Here, $f_{\mathrm{BBD}}$ is the density for the budding-birth-death process.

Next take the natural dual to Eq. \ref{ParPar}, the spontaneous-birth-assassination process
\begin{equation}
\varnothing \hh \substack{\mu_p \\  \longrightarrow} \hh A_p,\hl
A_p + A_q \hh \substack{\beta_{pq} \\ \longrightarrow} \hh A_p,
\end{equation}
which can be described by evolution operator
\begin{equation}
\mathcal{L}^\dag = \int_\Upsilon \rmd p \hs \mu_p(\psi_p^\dag-1)
 +\iint_{\Upsilon^2} \rmd p \hs\rmd q \hs \beta_{pq}(\psi_p^\dag\psi_p\psi_q -\psi_p^\dag\psi_q^\dag\psi_p\psi_q).
\end{equation}
This also results in a state evolution equation $\ket{\Psi_t} = \rme^{\mathcal{L}^\dag t}\ket{\Psi_0}$, where we have initial state $\ket{\Psi_0} = \ket{{\bf q}_n}$ and current state $\ket{\Psi_t} = \int_{\hat\Upsilon} \rmd \hat{\bf r} \hs f_{\mathrm{SBA}}(\hat{\bf r})\ket{\hat{\bf r}}$. Here, $f_{\mathrm{SBA}}$ is the density for the spontaneous-birth-assassination process.

Now these operators are connected by $\mathcal{L} = \mathcal{L}'+V$ where we have self adjoint operator
\begin{equation}
V = \int_\Upsilon \rmd p \hs \mu_p(\psi_p^\dag\psi_p-1)
 +\iint_{\Upsilon^2} \rmd p \hs\rmd q \hs \beta_{pq}(\psi_p^\dag\psi_p -\psi_p^\dag\psi_q^\dag\psi_p\psi_q).
\label{ImpAss}
\end{equation}

Note that we have the implicit assumption that $\int_\Upsilon \rmd p\hs \mu_p< \infty$. This will ensure the total spontaneous birthrate is finite and no population explosion occurs.

Then to construct a duality we use the function $C({\bf p}_m,{\bf q}_n)=\braket{{\bf p}_m|\rme^{\mathcal{L}^\dag t}|{\bf q}_n}$. On the one hand, using Eq. \ref{ChiDefRecov}, we have
\begin{equation}
C({\bf p}_m,{\bf q}_n) = \braket{{\bf p}_m|\Psi_t}
= \braket{{\bf p}_m|\int_{\hat{\Upsilon}} \rmd \hat{\bf r} \hs f_{\mathrm{SBA}}(\hat{\bf r})|\hat{\bf r}}
=f_{\mathrm{SBA}}({\bf p}_m),
\end{equation}
which is just the density at ${\bf p}_m$ for the spontaneous-birth-assassination particle process starting at ${\bf q}_n$. More explicitly we write $f_{\mathrm{SBA}}({\bf p}_m)=f_{\mathrm{SBA}}\left(\hat{\bf r}(t)={\bf p}_m|\hat{\bf r}(0)={\bf q}_n\right)$.

Conversely, following the approach to derive Eq. \ref{PINC}, a Feynman-Kac expectation of the following form is obtained, where
\begin{equation}
C({\bf p}_m,{\bf q}_n) = \braket{{\bf q}_n|\rme^{\mathcal{L}' t+Vt}|{\bf p}_m} = \int_{\hat{\bf r}(0)={\bf p}_m}^{\hat{\bf r}(t)={\bf q}_n} \mathcal{D}\hat{\bf r} \hs P[\hat{\bf r}] \rme^{\int_0^t \rmd s\hs V(\hat{\bf r}(s))},
\label{Cdefpi}
\end{equation}
and
\begin{equation}
V(\hat{\bf r}) = \left(\sum_{i=1}^{|\hat{\bf r}|}\mu_{\hat{\bf r}_i}-\int_\Upsilon \rmd p \hs \mu_p\right)-
\left(\sum_{\substack{i,j=1\\i \ne j}}^{|\hat{\bf r}|}\beta_{\hat{\bf r}_i\hat{\bf r}_i}-\sum_{i=1}^{|\hat{\bf r}|}\int_\Upsilon \rmd p \hs \beta_{\hat{\bf r}_ip} \right)
\end{equation}
Note that the first term is the variation in death rate, the second term is the covariation in birth rate, with $V$ representing variation of population decline for the budding-birth-death process. 

Now, the path integral in Eq. \ref{Cdefpi} is a sum of paths for vectors $\hat{\bf r}(s), s \in [0,t]$ (arising from the budding-birth-death process) with a specified start and end vector. The main difference from the derivation of Eq. \ref{PINC} is that the ket $\ket{z}$ is replaced with $\ket{{\bf q}_n}$ resulting in the extra boundary condition $\hat{\bf r}(t)={\bf q}_n$. A similar construction gives $\braket{{\bf q}_n|\rme^{\mathcal{L}'t}|{\bf p}_m}=f_{BBD}({\bf q}_n) = \int_{\hat{\bf r}(0)={\bf p}_m}^{\hat{\bf r}(t)={\bf q}_n} \mathcal{D}\hat{\bf r} \hs P[\hat{\bf r}]$, where we write $f_{BBD}({\bf q}_n)$ for the density at ${\bf q}_n$ of the budding-birth-death process starting at ${\bf p}_m$. Note in particular that the sum of the probability density functional $P[\hat{\bf r}]$ over all paths $\hat{\bf r}$ is restricted by the paths endpoints, and does not sum to unity in this case. Then normalising correctly, we obtain an expectation over paths with fixed termini,
\begin{equation}
\mathbb{E}_{BBD}\left(\exp\left\{\int_0^t \rmd s\hs V(\hat{\bf r}(s))\right\}\left|\substack{\hat{\bf r}(t)={\bf q}_n,\\ \hat{\bf r}(0)={\bf p}_m}\right.\right) = \frac{\int_{\hat{\bf r}(0)={\bf p}_m}^{\hat{\bf r}(t)={\bf q}_n} \mathcal{D}\hat{\bf r} \hs P[\hat{\bf r}] \rme^{\int_0^t \rmd s\hs V(\hat{\bf r}(s))}}{\int_{\hat{\bf r}(0)={\bf p}_m}^{\hat{\bf r}(t)={\bf q}_n} \mathcal{D}\hat{\bf r} \hs P[\hat{\bf r}]},
\end{equation}
resulting in the following duality, with a Feynman-Kac term written as the ratio of reciprocal densities,
\begin{equation}
\mathbb{E}_{BBD}\left(\exp\left\{\int_0^t \rmd s\hs V(\hat{\bf r}(s))\right\}\left|\substack{\hat{\bf r}(t)={\bf q}_n,\\ \hat{\bf r}(0)={\bf p}_m}\right.\right) = \frac{f_{SBA}(\hat{\bf r}(t)={\bf p}_m|\hat{\bf r}(0)={\bf q}_n)} {f_{BBD}(\hat{\bf r}(t)={\bf q}_n|\hat{\bf r}(0)={\bf p}_m)}.
\end{equation}

In general for particle models, although the condition $\bra{1}\mathcal{L} = 0$ is satisfied, the dual condition $\bra{1}\mathcal{L}^\dag=0$ is not and a Feynman-Kac correction will be needed to establish duality. However, dualities between pairs of particle models can be constructed in the manner above. 


\subsection{Diffusion Models}

We consider two cases, depending on whether we choose cylindrical or $Q$-Brownian motion. 

Firstly, consider the diffusion given by the following stochastic PDE, where $\rmd W_t$ is cylindrical Brownian motion in $H$, and the process $X_t \in H$ has some initial value $X_0=x$,
\begin{equation}
\rmd X_t(p) = \left(\nabla_p^2X_t(p)-X_t(p)+R_pX_t(p)^2\right)\rmd t + \omega_pX_t(p)\rmd W_t(p).
\label{SDSPDE}
\end{equation}
Thus we have two terms from the cable equation, a geometric noise term and a quadratic drift term. Then this process has a corresponding Liouvillian operator of the form
\begin{equation}
\mathcal{L}^\dag = \int_\Upsilon \rmd p \hs \psi_p^\dag(\nabla_p^2-1)\psi_p
+\int_\Upsilon \rmd p \hs R_p\psi_p^\dag\psi_p^2 
+\int_\Upsilon \rmd p \hs \frac{\omega_p^2}{2}(\psi_p^\dag)^2\psi_p^2.
\end{equation}
That is, we have evolution $\ket{\Psi_t} = \int \mathcal{D}z\hs P_X(z,t)\ket{z} = \rme^{\mathcal{L}^\dag t}\ket{\Psi_0} = \rme^{\mathcal{L}^\dag t}\ket{x}$ of initial state $\ket{x}$.

Next the operator $\mathcal{L}$ will be used to generate a dual process. The first term in $\mathcal{L}^\dag$ is self adjoint so results in identical drift terms in the corresponding stochastic PDE for $\mathcal{L}$. For the middle term of $\mathcal{L}$, we get $\int_\Upsilon \rmd p \hs R(p)(\psi_p^\dag)^2\psi_p$ which contributes cylindrical Brownian motion rather than a drift term. Combining with the third (self adjoint) term results in the following process $Y_t$, initiated with some function $Y_0=y$,
\begin{equation}
\rmd Y_t(p) = \left(\nabla_p^2Y_t(p)-Y_t(p)\right)\rmd t + \sqrt{Y_t(p)(2R_p+\omega_p^2Y_t(p))}\rmd W_t(p).
\end{equation}
Thus we have evolution $\ket{\chi_t} = \int \mathcal{D}z\hs P_Y(z,t)\ket{z} = \rme^{\mathcal{L} t}\ket{\chi_0} = \rme^{\mathcal{L} t}\ket{y}$ of initial state $\ket{y}$.

Then, to construct duality, we consider the function $C(y,x) = \braket{y|\rme^{\mathcal{L}^\dag t}|x}$ formed from the braket of two coherent states formed from the functions $x$ and $y$. Then recalling the product of two coherent states takes the form $\braket{x|y}=\rme^{\int_\Upsilon \rmd p \hs xy}$ we find, in much the same way as previous sections, that
\begin{equation}
C(y,x) = \mathbb{E}_X\left(\rme^{\int_\Upsilon \rmd p \hs X_t(p)y(p)}\right) = \mathbb{E}_Y\left(\rme^{\int_\Upsilon \rmd p \hs x(p)Y_t(p)}\right).
\end{equation}

If one attempts similar things with a $Q$-Brownian process, things become a little more awkward. Take for example,
\begin{equation}
\rmd X_t(p) = \left(\nabla_p^2X_t(p)-X_t(p)+X_t(p)\int_\Upsilon\rmd q\hs R_{pq}X_t(q)\right)\rmd t + X_t(p)\rmd W_t(p).
\end{equation}

This process generalizes the process above, with the choice $R_{pq}=R_p\delta(p-q)$ recovering Eq. \ref{SDSPDE}. The process has a Liouvillian of the form,
\begin{equation}
\mathcal{L}^\dag = \int_\Upsilon \rmd p \hs \psi_p^\dag(\nabla_p^2-1)\psi_p
+\iint_{\Upsilon^2} \rmd p \hs\rmd q\hs R_{pq}\psi_p^\dag\psi_p\psi_q 
+\iint_{\Upsilon^2} \rmd p \hs\rmd q\hs \frac{\Omega_{pq}}{2}\psi_p^\dag\psi_q^\dag\psi_p\psi_q,
\end{equation}
where $\Omega_{pq} = \sum_k\lambda_k\xi_k(p)\xi(q)$. 

Now the first and third terms are self adjoint. Thus for $R_{pq}=0$ we can use the construction above to get a self dual process with duality function $\rme^{\int_\Upsilon \rmd p \hs xy}$. However, the lack of symmetry between $p$ and $q$ in the second term $\iint_{\Upsilon^2} \rmd p \hs\rmd q\hs R_{pq}\psi_p^\dag \psi_p\psi_q$ means that the adjoint does not correspond to a $Q$-Brownian motion term. It is also not a function of density operator $\psi_p^\dag\psi_p$ so cannot be transformed into a Feynman-Kac term as in previous sections, and other ideas are needed to find dualities.


\section{Conclusions}

The methods described above allow Doi field theoretic methods to establish dualities between particle processes and diffusion in Hilbert spaces, generalizing the methods of \cite{Ohkubo2013}. In terms of models, we find that a particle process of the form $A_p \longrightarrow\dots$ will give rise to a drift term in the dual diffusion process, and a process of the form $A_p+A_q \longrightarrow\dots$ will give rise to a Brownian motion term. Furthermore a local process such as $A_p+A_p \longrightarrow\dots$ will correspond to cylindrical Brownian motion, whereas a non-local process such as $A_p+A_q \longrightarrow\dots$ will correspond to $Q$-Brownian motion. This naturally raises the question of whether there are dualities of interest for more complex particle processes such as $A_p+A_q+A_r \longrightarrow \dots\hh$. The corresponding Fokker-Planck equation will have third order differentials, suggesting if a dual stochastic process exists, it is not Brownian motion. Indeed any such dual `process' may not be stochastic. We have seen in some cases that duality is with a deterministic process. It would seem feasible that if the requirement is loosened so that the target is just a signed measure (rather than a positive probability measure), for example, more `dualities' may be possible.

The Doi methods described are applicable when the drift and noise terms are polynomial in nature. For example, the stochastic cable equation that was analysed contained a linear drift term. More detailed models \cite{Mckean1970} suggest a cubic model may be more precise. This will be amenable to the kind of analyses we have employed, although this will entail a perturbative path integral expansion that will contain internal nodes corresponding to the cubic terms, which will paint a more complicated picture than that seen in Fig. \ref{Feynman}. Whether these methods can be adapted to more general non-polynomial forms is an open problem, the solution of which would certainly increase its utility. In some cases non-polynomial systems can be analysed, but this currently relies on transforming the system to a polynomial form (see conclusions in \cite{Ohkubo2013}, for example). 

For most processes discussed, the operator $B(x)$ has taken constant, or local, somewhat uninteresting forms $B(x)(p)=b(x(p))$, where the function $B(x)$ at $p$ only depends on the function $x$ at $p$. The jump-diffusion process had a `jump' operator $\mathcal{L} = \iint \rmd p\hs\rmd q \hs R_{pq}(\psi_q^\dag\psi_p-\psi_p^\dag\psi_p)$, where $R_{pq}$ represented the rate a particle at position $p$ hops to $q$. Under duality, this translated to the drift term $B(x)(p) = \int \rmd q R_{pq}(x(q)-x(p))$ in the corresponding stochastic PDE, resulting in a function $B(x)(p)$ that depends upon the entire function $x$. A fuller exploration of the range of possible operators $B(x)$ that arise from Liouvillian particle operator counterparts would certainly be of interest. These questions also apply to possibilities that will arise by considering more than one species of particle, and spatial processes other than diffusion; the convection terms in age structured systems, for example, may offer alternative features of interest \cite{Greenman1}, \cite{Greenman2}. In some cases (e.g Eq. \ref{ImpAss}) we have seen restriction on coefficients to ensure the number of particles in the systems is finite. However, there are well characterized techniques for countably infinite systems of particles \cite{Matthes1978,Albeverio1998}, and extending these methods for these cases would be useful.

The Doi functionality has enabled duality to be established on quite a wide scale. However, there are a host of other classes of duality functions. For example, the duality function $\mathbb{I}_{x \le y}$ exists for a wide class of Feller processes, which is provable by other means \cite{Liggett2012}. Whether other field theoretic approaches can connect these is an open problem.

Dealing with infinite dimensional diffusion is fraught with technical difficulties, and conversely, path integrals are notorious for their need of greater rigour. However, these methods do offer a means of exploring dualities between stochastic processes that can then be investigated with alternative methods.

\bibliographystyle{abbrv}
\bibliography{DoiDuality}

\end{document}